\documentclass[journal]{IEEEtran}
\usepackage{amsmath}
\usepackage{amssymb}
\usepackage{amsthm}
\usepackage{bm}
\usepackage{lipsum}
\usepackage{cuted}
\pdfoutput=1
\usepackage[dvips]{graphicx}
\usepackage{algorithm}
\usepackage{algorithmic}
\usepackage{slashbox}
\usepackage[labelformat=simple]{subcaption}

\usepackage{hyperref} 
\usepackage[numbers,sort,compress]{natbib} 

\usepackage{color}
\usepackage{slashbox}
\usepackage{multirow}
\usepackage{multicol}

\usepackage[normalem]{ulem}

\begin{document}
\title{A Dual-Cluster-Head Based Medium Access Control for Large-Scale UAV Ad-Hoc Networks}
\author{Xinru Zhao, Zhiqing Wei, Yingying Zou, Hao Ma, Yanpeng Cui, Zhiyong Feng}
\maketitle
\begin{abstract}
 Unmanned Aerial Vehicle (UAV) ad hoc network has achieved significant growth for its flexibility, extensibility, and high deployability in recent years. The application of clustering scheme for UAV ad hoc network is imperative to enhance the performance of throughput and energy efficiency. In conventional clustering scheme, a single cluster head (CH) is always assigned in each cluster. However, this method has some weaknesses such as overload and premature death of CH when the number of UAVs increased. In order to solve this problem, we propose a dual-cluster-head based medium access control (DCHMAC) scheme for large-scale UAV networks. In DCHMAC, two CHs are elected to manage resource allocation and data forwarding cooperatively. Specifically, two CHs work on different channels. One of CH is used for intra-cluster communication and the other one is for inter-cluster communication. A Markov chain model is developed to analyse the throughput of the network. Simulation result shows that compared with FM-MAC (flying ad hoc networks multi-channel MAC,FM-MAC), DCHMAC improves the throughput by approximately 20\%-50\% and prolongs the network lifetime by approximately 40\%.

\end{abstract}

\section{INTRODUCTION}
\label{Introduction}
Unmanned Aerial Vehicles (UAVs) are characteristic of mobility, flexibility and low cost \cite{1}, which is widely applied in civil and military fields, particularly in the scenarios such as environment and natural disaster monitoring, communications in remote areas, emergency search and rescue \cite{2,3,4,5,6,26}. Besides, the UAV-assisted scenario is also promising in maritime communication coverage enhancement \cite{27}. Compared with single UAV, UAV ad hoc network, also known as flying ad hoc network (FANET), is more self-organizing and distributed, which is more suitable for these scenarios. Despite its high scalability and versatility over single UAV, the deployment of FANET suffers from several major challenges, such as management of a large number of UAVs, dynamic mobility, and resource constraints. In overcoming these challenges, the design of efficient medium access control (MAC) strategy is indispensable for large-scale FANETs, as it is a prerequisite to ensure communication efficiency and network lifetime \cite{7}.

Due to the lack of global knowledge for dynamic topology and decentralized channel contention, the conventional flat network structure is inadequate in large-scale network as FANET. Thus, the implementation of clustering is requisite to address this issue \cite{8,9}. In the clustering scheme, the network is divided into a number of clusters, with each cluster contains a cluster head (CH) and several cluster members (CMs). CH is responsible for managing all the CMs, resources, and data forwarding. CMs are used for specific and detailed execution of tasks. However, because of the high dynamic topology and high throughput requirements, it brings out an interesting and contradictory expectation of MAC mechanism for nodes in the cluster. On the one hand, CMs desire for the simplicity of network structure and communication mechanism to achieve guileless communication procedure and reduce energy consumption. On the other hand, they require multi-dimensional resources to improve throughput and achieve load balancing when traffic is concurrent.

An optional solution to the above challenge is the method of adopting different access schemes for intra-cluster and inter-cluster communications \cite{11,12,13}. However, these schemes are all rely on single CH in each cluster, which will inevitably lead to the increasing overload and premature death of CH when the number of CMs increased. Dual cluster head (DCH) scheme provides a light-weighted solution to implement load balancing, which can be used as a complementary method to improve throughput and energy efficiency~\cite{10,14,15,16,17}. Nevertheless, research in terms of this kind of scheme is mainly focused on terrestrial scenarios such as vehicular ad hoc network (VANET), which is not suitable for FANET. Despite years of intensive research, the main limitation of most schemes stems from the cooperation between the management of numerous nodes and resource allocation, which hinders the application of these schemes in high dynamic UAV networks.
 
Motivated by the above considerations, this paper presents a dual-cluster-head based MAC (DCHMAC) scheme for cluster-based UAV ad hoc network. Specifically, there are three contributions of DCHMAC scheme:

1) A novel architecture for cluster-based UAV ad hoc network is considered, which improves throughput in contrast to the conventional cluster-based network with single CH. To the best of our knowledge, this is the first MAC scheme that uses a dual cluster head mechanism in the UAV ad hoc network.

2) The channels are divided according to different communication scenarios. Different scenarios apply different MAC schemes to balance the network workload through the cooperation of two CHs. The frame structures are also optimized to improve the adaption of the network to the high mobility of UAV.

3) Theoretical analysis and simulations are performed to further verify the effectiveness of the proposed scheme. The simulation results show that the proposed scheme can improve the throughput and lifetime of the network.

The rest of the paper is organized as follows. In Section~\ref{Related works}, the related works are analyzed. Section~\ref{MODEL} describes the system model of the UAV network. Section~\ref{ALGORITHM} provides the clustering algorithm. Section~\ref{DCHMACDESIGN} presents the DCHMAC design in detail, including the channel division method and frame structure design. The theoretical performance and simulation results are discussed in Section~\ref{performance analysis} and Section~\ref{simulation results}, respectively. Finally, we conclude the paper in Section~\ref{conclusion}.

\section{RALATED WORKS}
\label{Related works}
In order to improve the performance of cluster-based networks, many studies have focused on the method of adopting various access schemes for intra-cluster and inter-cluster communications. In \cite{11}, a cluster-based multichannel MAC mechanism based on time division multiple access (TDMA) is proposed for VANET. In the proposed mechanism, vehicles which move in the same direction are grouped into clusters, and CH is responsible for the transmission slots assignment. Each synchronization interval consists of a control channel interval (CCHI) and a service channel interval (SCHI) to improve the throughput of the network. To solve intra-cluster and inter-cluster transmission problems, a collision-free multichannel MAC (CCFM-MAC) protocol \cite{12} is proposed, in which CH allocates time slots based on the state and relative location of its CMs. For reducing the interference and hidden terminal problem in VANETs, a cluster-based interference-free MAC scheme with load aware for non-safety message communication (CMNSM) \cite{13} is presented, which is composed of inter-cluster multichannel allocation and intra-cluster dynamic TDMA frame allocation.

The above schemes are all under the consideration of single CH within each cluster, which will inevitably lead to the increasing schedule complexity and overload of CH when the number of UAVs increased. Aiming at this problem, the authors in \cite{14} propose a double cluster head scheme considering the communication scenario of base stations (BSs). In this scheme, CM sends data packets to two CHs when away from the BS, and the packets will be carried until CH is within the range of any BS. The protocol proposed in \cite{15} makes some improvements. When CM sends data to the node outside the cluster, it first sends the data to two cluster heads. If RSU is available, the primary CH (PCH) is responsible for forwarding. Otherwise, the secondary CH (SCH) broadcasts to the neighbor cluster head, in this case, all the CHs broadcast to the neighbor CHs until packets are successfully transmitted to the destination cluster. In \cite{16}, a double-head clustering algorithm is developed for VANET. When PCH loses links with its CMs, SCH will relay the data packets it received from its neighbor to PCH. Study in \cite{17} presented a trust-aware double CH routing scheme. When PCH fails to work normally, SCH replaces PCH to maintain the reliability of cluster head.

The above schemes all rely on fixed infrastructure. Besides, the topology and the link status of FANET are more complicated than VANET. Therefore, these schemes are sometimes inapplicable in FANET. There is limited research on the dual cluster head scheme in the cluster-based UAV networks \cite{18}. For FANET, the MAC protocols are categorized into two types, i.e. contention-based and contention-free. These MAC protocols can be integrated with directional antennas and omni-directional antennas to exploit channel efficiency and spatial utilization. Sensing multiple access with collision avoidance (CSMA/CA) is the most often used scheme in contention-based methods, TDMA is the most typical scheme in contention-free methods \cite{19}. For example, the authors in \cite{20} proposed a CSMA/CA based MAC protocol (FMAC) for FANET under density-varying flocking scenarios. However, only the broadcast information service is discussed in this scheme, which limited its application in the directional communication scenario. In \cite{21}, a multi-cluster-based FANETs for disaster situation is presented, in which time slots for intra-cluster and inter-cluster are assigned by CH. UAVs make use of adaptive transmission power for communication. In \cite{22}, authors proposed a self-organized slot access process of neighboring cooperation to avoid unavailable topology information and information exchange. These two mechanisms can be suitable in a small-scale UAV network. When the number of slots and UAVs increased, the mechanisms would be intricate. In \cite{7}, authors proposed a multi-channel MAC protocol (FA-MMAC) in FANET by adopting retro-directive array antennas. The FA-MMAC improves the system performance by employing multi-channel and retro-directive array antennas. Adjacent UAVs can realize simultaneous transmission over the same channel without interference. However, implementing this kind of retro-directive array antennas is impractical for numerous UAVs considering hardware storage limitation and cost. By combining the advantages of multi-channel and directional antenna, authors in \cite{23} propose a FM-MAC protocol to improve system performances. In this scheme, mobile prediction and preemption mechanisms are adopted to achieve high throughput and low delay. Control packets for safety in CCHI are transmitted in omni-directional way, whereas data packets are transmitted in SCHI in the desired direction. The mobility information is periodically broadcasted to guarantee flight safety. However, the length of CCHI and SCHI is immutable, which is inadaptable for different scale of the UAV network.

\section{SYSTEM MODEL}
	\label{MODEL}
The UAV ad hoc network consists of multiple clusters. Intra-cluster communications of adjacent clusters are organized in different channels to avoid interference. Each cluster selects a PCH and a SCH. PCH is used for inter-cluster communication and SCH is for intra-cluster communication. Each cluster is composed of numerous UAVs, and the UAVs in the cluster form a muti-hop network. Other assumptions are as follows:
	
1) Each UAV is equipped with a GPS and two transceivers;
	
2) Control message is transmitted by omnidirectional broadcasting and data message is transmitted in oriented way.
	
3) PCH has two transmission power levels. The larger one empowers PCH reachable to its neighbor PCHs, and the smaller one is used for omnidirectional broadcasting in the cluster.
	
4) CM also has two transmission power levels. The larger one is used to send messages to its SCH with orientation, and the smaller one is used for omnidirectional broadcast within the range of one-hop distance.
	
\begin{figure}[htbp]
		\centering
		\includegraphics[width=0.45\textwidth]{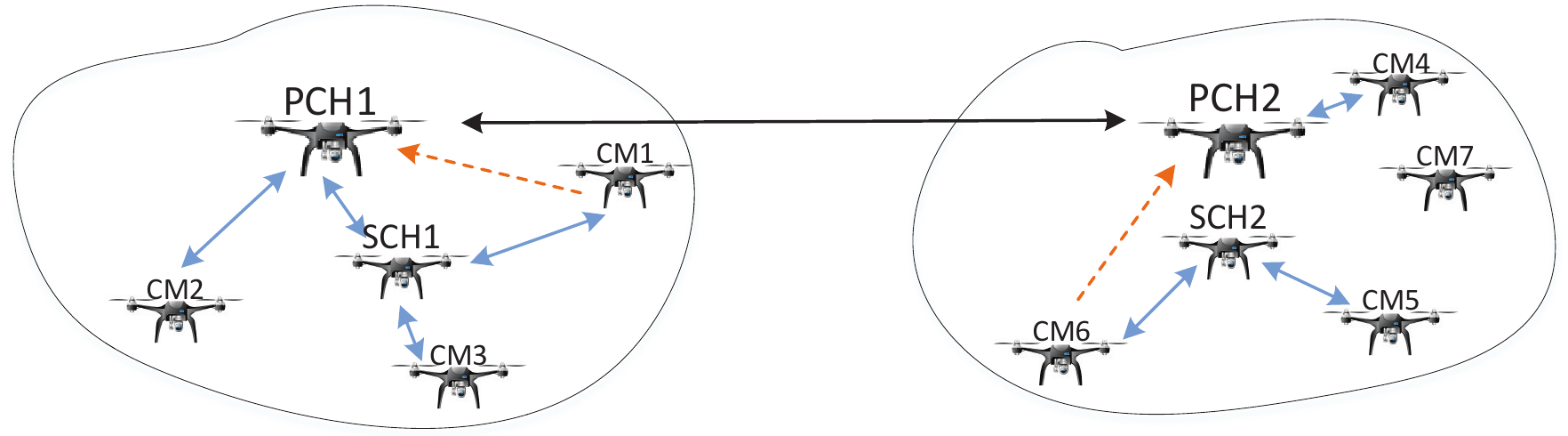}
		\caption{Clustering based UAV network.}
		\label{fig1}
	\end{figure}

Figure~\ref{fig1} shows the structure of the UAV network. In the DCHMAC scheme, the conventional data forwarding task of CH is now jointly undertaken by PCH and SCH. PCH is mainly in charge of inter-cluster data forwarding and resource reservation. SCH is responsible for intra-cluster data forwarding between CMs whose distance are beyond one-hop range. The communication between PCH and SCH is mainly about resource allocation and notification. This kind of collective teamwork between PCH and SCH improves not only the network lifetime but also data stream division. As Figure~\ref{fig1} shows, the solid line represents data communication
and the dotted line represents resource reservation. If CM1 wants to communicate with CM3, it sends the reservation request to PCH1. After the reservation is successful, the message is forwarded to CM3 through SCH1. Besides, if CM2 wants to communicate to CM4, the successful message forwarding procedure will be CM2 to PCH1, PCH1 to PCH2 and finally PCH2 to CM4.
 
\section{CLUSTERING ALGORITHM}
\label{ALGORITHM}
Our clustering algorithm mainly compose of three parts: cluster formation, PCH and SCH selection, and cluster maintenance. 
	
1) Cluster formation
	
Assume that the total number of UAVs is $N$ and the number of clusters is $M$. Each UAV has a unique ID in the whole network and a unique CID (cluster ID, CID) in the cluster. The CID of CM is greater than 2.
	
	a) Initiation  
	
Each UAV generates a random value $P$ between 0 and 1. The node whose value satisfies Eq.~\eqref{eq1} automatically becomes TCH (temporary CH, TCH).
	\begin{equation}
		p = rand(0,1) \leqslant \frac{M}{N}.
		\label{eq1}
	\end{equation}
	
	b) Cluster formation
	
\begin{figure}[htbp]
		\centering
		\subfloat[HELLO packet]{\label{rjc}
			\includegraphics[width=0.45\textwidth]{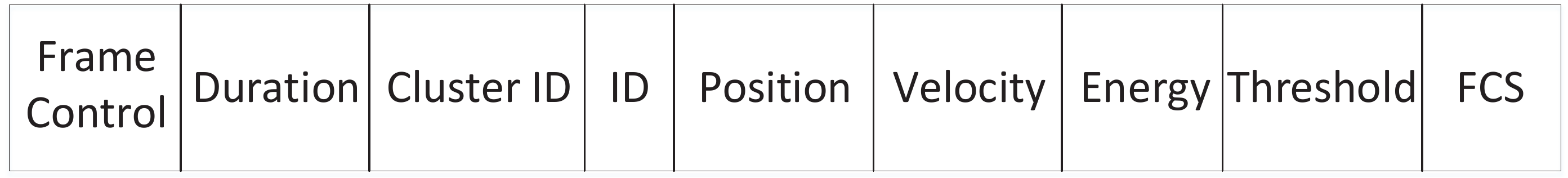}
		}\\
		\subfloat[RJC packet]{\label{ajc}
			\includegraphics[width=0.45\textwidth]{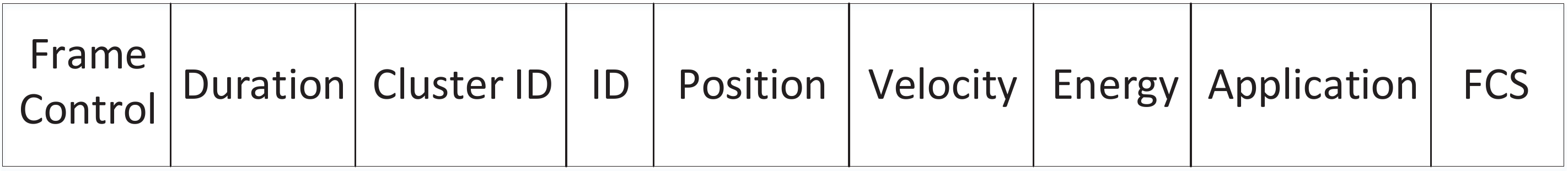}
		}\\
			\subfloat[DCH frame structure]{\label{DCH}
		\includegraphics[width=0.45\textwidth]{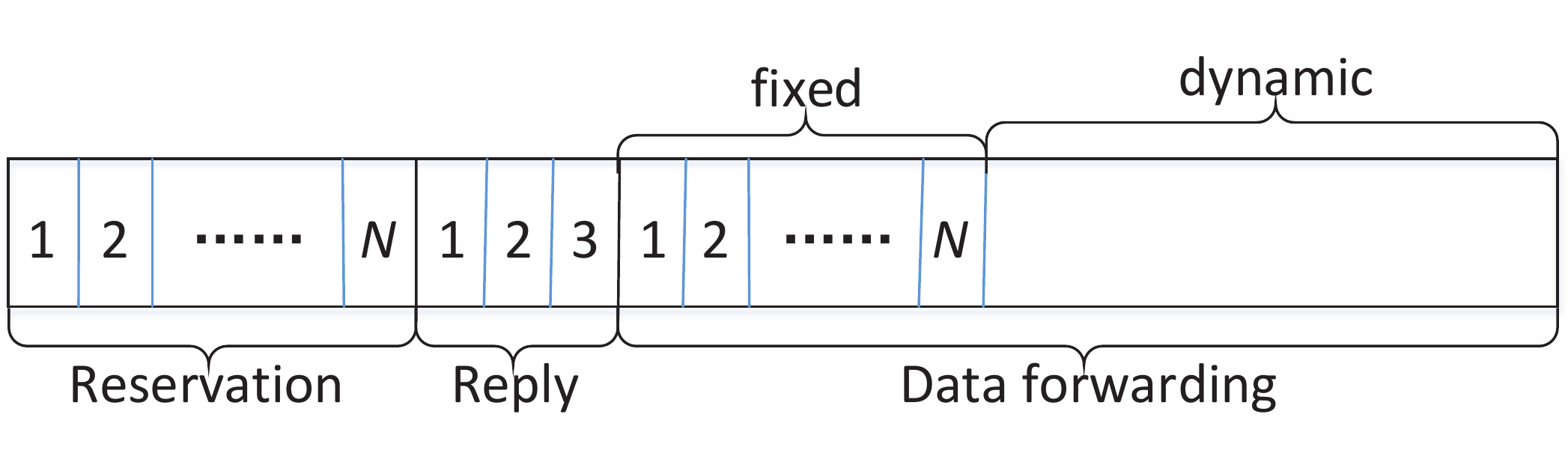}
	}\\
		\subfloat[Reservation packet]{\label{reservation}
			\includegraphics[width=0.45\textwidth]{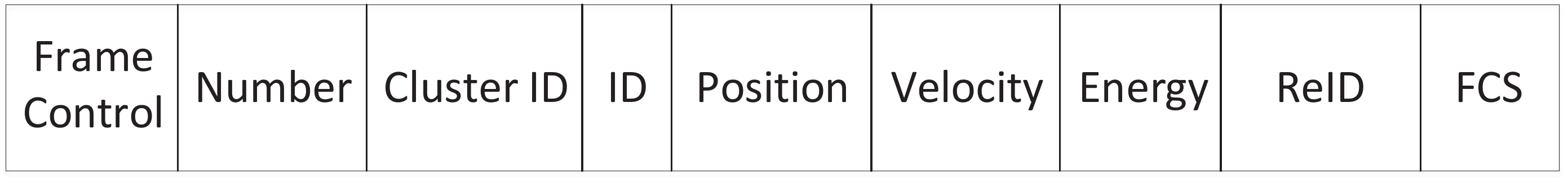}
		}
	\caption{Frame structure}
		\label{fig2}
	\end{figure}

Each TCH broadcasts a HELLO message containing the CID, ID, location, speed, remaining energy and the threshold of the number of CMs. Figure 2\subref{rjc} shows the frame format of HELLO packet. The CID is same as the ID of the PCH, and it is the value of TCH’s ID at this stage. Each TCH updates a CNAV (cluster network allocation vector, CNAV) list by receiving HELLO packets containing the informantion of adjacent clusters.

Each node that is not a TCH keeps listening. If it receives a HELLO message, the node will broadcast a RJC ( request to join the cluster, RJC) packet containing its own ID, location, speed, remaining energy and application information. Figure 2\subref{ajc} shows the frame format of RJC packet. The minimum length of the Application is 1 bit. 0 and 1 represents joining and quiting the cluster respectively. If TCH agrees the node to join the cluster, it will add the information of the node to its NAV (network allocation vector, NAV) list.
	
After receiving RJC packets, TCH needs to decide whether to accept the application, according to the residual number of idle CIDs. Then TCH broadcasts an AJC (agree to join the cluster, AJC) packet, in which packet includes the information of CID (namely its own ID), the IDs that are accepted to join the cluster and corresponding CIDs assigned to them, and TCH's location, speed, remaining energy, as well as the remaining number of idle CIDs. 
	
Nodes acquire the information of IDs and CIDs of other nodes from AJC packets, and the information is recorded in the local NAV list. If a node haven't received any HELLO message for a long time, it is considered that the node is far away from its adjacent clusters. In this case, the node declares itself TCH and broadcasts a HELLO message. If a node receives more than one HELLO message, it will send RJC packet to its nearest TCH.
	
2) PCH and SCH selection
	
To improve the stability and the lifetime of UAV network, PCH and SCH are elected considering the position, speed, direction, and residual energy of all UAVs in the cluster. Since TCH has the information of all the CMs, it will specify PCH and SCH according to the value of CFCH (the chance for CH, CFCH) of each node. Assume that the position of node $i$ is ($x_i$,$y_i$), its residual energy is $P_i$, then its CFCH value is
	\begin{equation}
		CFCH_i = w_1P_i + w_2\frac{1}{D_i} +w_3\frac{1}{T_i},
		\label{eq2}
	\end{equation}
in which, $w_1$, $w_2$ and $w_3$ are the weighting factors, satisfying $w_1+w_2+w_3 = 1$. $D_i$ is the differential value between the velocity of node $i$ and the average velocity of other nodes in the cluster. $T_i$ is the Euclidean distance between the position of node $i$ and the average position of other nodes in the cluster.	
The two nodes with the largest CFCH value will be appointed by TCH as PCH and SCH respectively, and their CIDs will be automatically set to 1 and 2. After the election, TCH broadcasts the information of PCH and SCH, including ID, position and speed. Hence all the CMs in the cluster and the CHs of adjacent clusters will be informed of the status information of the PCH and SCH accordingly.
	
3) Cluster maintenance
	
When a new node joins a cluster, it will broadcast a RJC packet to its neighbor nodes. Since the new node doesn’t know the PCH’s position, the CM with the smallest CID will forward the RJC packet. The number of CMs in the cluster will change dynamically due to the mobility of UAVs. If PCH does not receive any messages from a CM for four consecutive times, it will record the CID of the node as idle and then allocate the CID to a new node. When the number of departures in the cluster exceeds the threshold in a short time, it is considered that the network topology and cluster members are changed greatly, then PCH will broadcast a cluster dissolution message. If the remaining energy of the SCH is insufficient, a new SCH will be assigned.   
\section{DCHMAC DESIGN}
	\label{DCHMACDESIGN}
In this section, the DCHMAC scheme is described, including channel division method and frame structure design.
\subsection{Channel division}
\begin{figure}[htbp]
		\centering
		\includegraphics[width=0.45\textwidth]{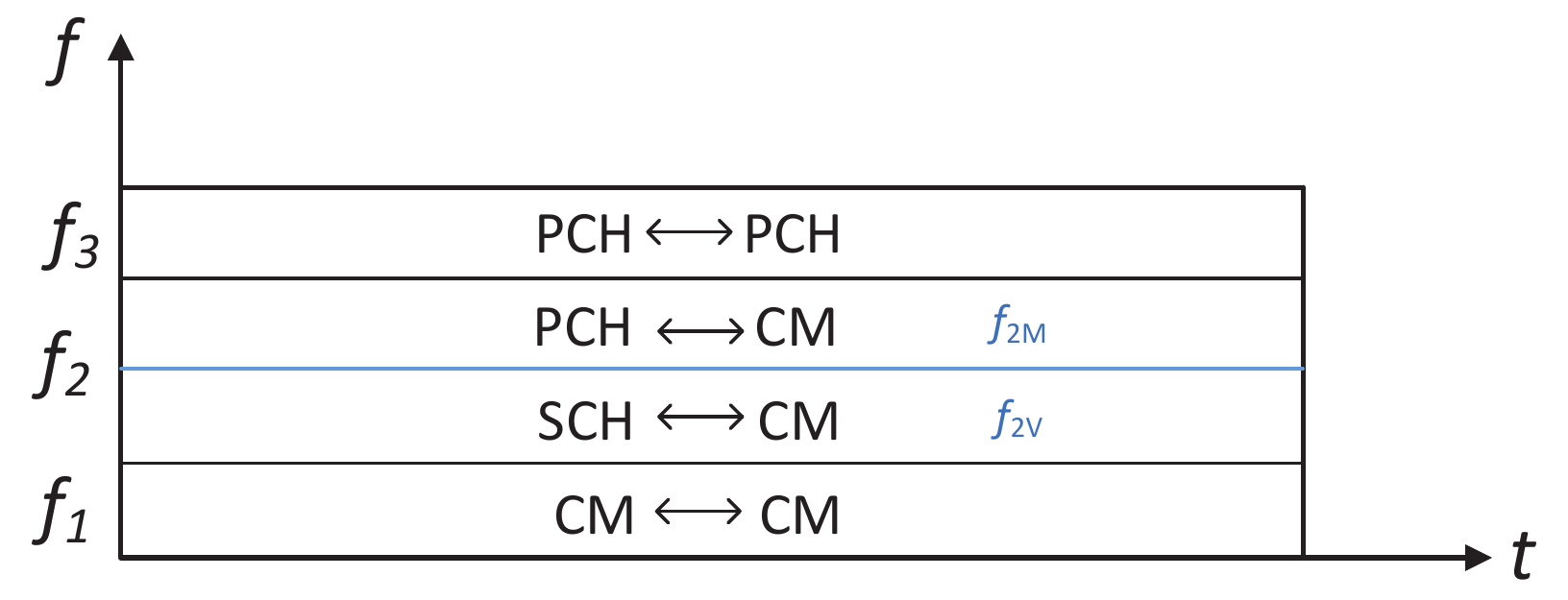}
		\caption{Channel division.}
		\label{fig3}
	\end{figure}

As Figure~\ref{fig3} shows, the whole valuable channel is divided into three main segments, represented by $f_1$, $f_2$, and $f_3$ respectively. The communication scenarios of the channel segments are summarized as follows:
	
	1) $f_1$ is used for all nodes during cluster formation stage and for one-hop broadcast of CMs after cluster formation stage. Since CM’s broadcast range is relatively small, this segment can be used in parallel in different clusters or in different regions with a long distance in a cluster.  
	
	2) $f_2$ is used for intra-cluster communication. The value of $f_2$ between adjacent clusters is different. Each $f_2$ is divided into two sub-channels, one is $f_{2M}$ for PCH and CM communication, and the other is $f_{2V}$ for SCH and CM communication.  
	
	3) $f_3$ is used for inter-cluster communication of PCHs. 

Two transceivers of nodes work on different channels, which is described as follows:

a) For PCH, one transceiver works in $f_3$ for inter-cluster communication, the other one works in the local $f_{2M}$ for intra-cluster communication.

b) For SCH, one transceiver works in $f_{2V}$, which is used for long distance data forwarding between CMs in the cluster, the other one works in $f_1$ for broadcasting.

c) For CM, one transceiver works in $f_{2M}$ for inter-cluster data forwarding or in $f_{2V}$ for intra-cluster communication, the other one works in $f_1$ for one-hop broadcast.

\subsection{Frame structure design}
The design of DCHMAC adopts the hybrid of TDMA and CSMA. The inter-cluster communication and the one-hop communication use CSMA scheme. The nodes whose distances are beyond one-hop range adopt TDMA. The communication frame structure of DCHMAC consist of three stages, namely the reservation period, the reply period and the data transmission period, as Figure 2\subref{DCH} shows.  
	
1) Reservation period
	
CM reserves time slots during this period for two purposes: data transmission and periodically informing that it is still in the cluster. The length of the reservation period is fixed with $N$ time slots, where $N$ is the maximum number of nodes that a cluster can accommodate. Each node knows when to send the reservation message according to the value of CID. Since the CID of PCH is 1 and that of SCH is 2, the first time slot of this period is fixed to SCH. The frame format of the reservation message is shown in Figure 2\subref{reservation}, where Number is the number of time slots reserved by CM, including the number of dynamic time slots required for inter-cluster communication and intra-cluster communication. ReID is the ID of the corresponding communication destination node. 
	
	2) Reply period
	
There are three time slots in the reply period. In the first time slot, PCH communicates with SCH about the reservation message for intra-cluster communication, which includes ID pairs, the status information of sender and receiver, and the number of time slots required. In the second time slot, SCH replies to PCH on the ID pairs that are successfully reserved. In the third time slot, PCH broadcasts reply message in sub-channel $f_{2M}$, including the ID pairs, time slot allocation results, the status information of PCH and SCH. 
	
3) Data forwarding period
	
After receiving the broadcast message, each CM knows the time slot allocation result. If the reservation is successful, CM will send inter-cluster messages to PCH and/or send intra-cluster messages to SCH in the corresponding time slots. By consulting ReID, the destination node knows when to keep in sensing state. If it finds that it needs to receive the intra-cluster message and the inter-cluster message at the same time, another transceiver in $f_1$ will adjust to $f_{2V}$. Since CM knows the sending and receiving time slots of other nodes, if it finds that both transceivers of a one-hop neighbor are occupied by $f_2$ in a certain period, it will not broadcast message to this node during this period.   
	
The data transmission period of PCH is divided into fixed period and dynamic period. The length of the fixed period can be multiple time slots. The functions of fixed period are as follows:  
	
	a) It is used for the transmission of periodic messages or inter-cluster messages; 
	
	b) the message transmitted during this period carrys the state information of CM, so PCH can acquire CM's state twice in a frame, where another acquirement is in the reservation period. After two frames, PCH will know whether a node has already left the cluster.
	
	c) When a new node request to join the cluster, the CM who is responsible for forwarding the RJC packet will forward the packet to PCH in this fixed period. Then PCH broadcasts the AJC packet in the reply period at the next frame. So it takes no more than one frame for a node to join the cluster.
		
the transmission period of SCH and PCH are different. SCH is only responsible for intra-cluster forwarding, so there is no reservation period. In addition, as long as the distance between the sender and the receiver is beyond one-hop range in the cluster, the CM will send messages to SCH to forward it. This forwarding method can save the storage space of all CMs, because CM only needs to maintain the information of nodes within one-hop range and the status information of CHs. There is no need for a CM to know the position and speed of distant nodes in the cluster. This method is suitable for the large scale UAV network communication.
	
\section{PERFORMANCE ANALYSIS}
\label{performance analysis}
\begin{figure}[htbp]
		\centering
		\includegraphics[width=0.45\textwidth]{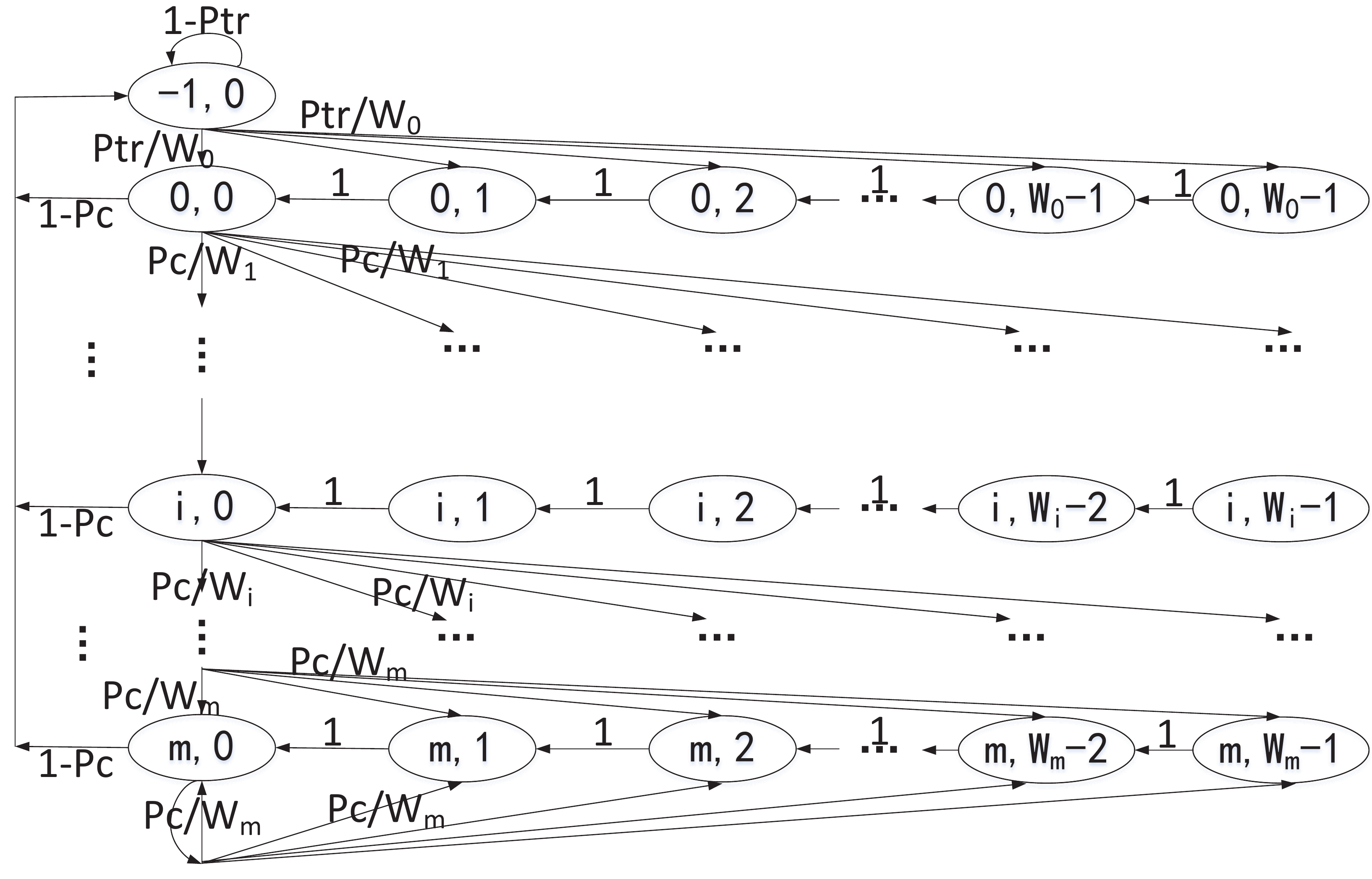}
		\caption{Markov chain model.}
		\label{fig4}
	\end{figure}

The two-dimensional Markov chain is adopted to model the backoff mechanism of the inter-cluster communication. As Figure~\ref{fig4} shows, we use $b_{i,j} = \lim_{t \to \infty}P\{u(t)=i,v(t)=j\}$ to denote the steady probability of every state, where $i$ is the backoff stage, $j$ is the size of the contention window. The maximum value of $j$ for $i$ is $W_i$, which satisfies $W_i=2^iW$, $W$ is a pre-set value. In the Markov chain, $\{(i,j) \mid i \in \{0,1,2,...,m\},j \in \{0,1,2,...,W_i-1\}\}$ denotes the steady state of the node. 
	The state (-1,0) denotes the state after a message is transmitted successfully. So the state transmission probability is 
 \begin{equation}\small
	\begin{aligned}
		\begin{array}{ll}
			P\{ 0,j| - 1,0\}  = \frac{{{P_{tr}}}}{{{W_0}}},&j \in [0,{W_0} - 1]\\
			P\{  - 1,0|i,0\}  = 1 - {P_c},&i \in [0,m],j \in [0,{W_0} - 1]\\
			P\{ i,j|i,j + 1\}  = 1,&i \in [0,m],j \in [0,{W_i} - 2]\\
			P\{ i,j|i - 1,0\}  = \frac{{{P_c}}}{{{W_i}}},&i \in [1,m],j \in [0,{W_i} - 1]\\
			P\{ m,j|m,0\}  = \frac{{{P_c}}}{{{W_m}}},&j \in [0,{W_m} - 1].
		\end{array}
	\end{aligned}
\end{equation}
the possibility of the steady state satisfies
\begin{equation}
	\begin{array}{l}
		{b_{ - 1,0}} = \frac{1}{{{P_{tr}}}}{b_{0,0}}\\
		{b_{ i- 1,0}}{P_c} = {b_{i,0}} \Rightarrow {b_{i,0}} = P_c^i{b_{0,0}}\\
		{b_{m - 1,0}}{P_c} + {b_{m,0}}{P_c} = {b_{m,0}} \Rightarrow {b_{m,0}} = \frac{{{P_c}^m}}{{1 - {P_c}}}{b_{0,0}}.
	\end{array}
\end{equation}
Then we have
\begin{equation}
	\begin {aligned}
{b_{i,j}} = \frac{{{W_i} - j}}{{{W_i}}}\left\{ \begin{array}{ll}
		(1 - {P_c})\sum\limits_{k = 0}^m {b_{k,0}}, & i = 0 \\
		{P_c}{b_{i - 1,0}}, &0 < i < m\\
		{P_c}({b_{m - 1,0}} + {b_{m,0}}), &i = m.
	\end{array} \right.
\end{aligned}
\end{equation}
$P_{tr}$ is the probability of inter-cluster communication between PCHs. $P_c$ and $P_b$ are the probability of collision and the probability of channel busy, respectively. Summarize all the possibilities of Figure~\ref{fig4} equals 1, which is
\begin{equation}
\sum\limits_{i = 0}^m {\sum\limits_{j = 0}^{{W_i} - 1} {{b_{i,j}}}  + {b_{ - 1,0}} = 1} ,
\end{equation}
so we have
\begin{equation}
	{b_{0,0}} = \frac{1}{{\frac{{(1 - 2{P_c})(W + 1) + {P_c}W(1 - {{(2{P_c})}^m})}}{{2(1 - 2{P_c})(1 - {P_c})}} + \frac{1}{{P{}_{tr}}}}},\\
\end{equation}
the possibility of transmission is expressed as Eq.~\eqref{eq7}.
\begin{figure*}[ht]
\begin{equation}
\begin{aligned}
	\tau  &= \sum\limits_{i = 0}^m {{b_{i,0}}}  = \frac{1}{{1 - {P_c}}}{b_{0,0}}{\kern 1pt} \\
	&= \frac{{2{P_{tr}}(1 - 2{P_c})}}{{((1 - 2{P_c})(W + 1) + {P_c}W(1 - {{(2{P_c})}^m})){P_{tr}} + 2(1 - 2{P_c})(1 - {P_c})}}.
	\label{eq7}
\end{aligned}
\end{equation}
\end{figure*}

A. The throughput of a single cluster

We assume that the total number of UAVs is $N$, the number of clusters is $M$, the average number of UAVs in each cluster is $P$, and the number of CMs in each cluster is $Q$. Considering the mobility of UAV, assuming that UAV’s average arrival rate to each cluster is $\lambda_1$, the density of UAV is $\rho_v$, and its relative speed compared with PCH is $v_r$. The maximum capacity of UAVs in a cluster is $X_M$, the broadcast range of PCH is $R_{trans}$. The ralative free-flow speed of CM to PCH is $v_f$, at which speed UAV congestion does not occur, then \cite{11}
\begin{equation}
\begin{aligned}
	\lambda_1  &= {\rho _v}{v_r}\\
	{\rho _v} &= {\textstyle{{{X_M}} \over {{R_{trans}}}}}(1 - \frac{{{v_r}}}{{{v_f}}}),
\end{aligned}
\end{equation}
thus the average nodes in a cluster is
\begin{equation}
E(X) = \lambda_1 {T_s},
\end{equation}
where $T_s$ is the average period of Markov state. The expectation of $N$ is
\begin{equation}
E[N] = ME(X),
\end{equation}
the relationship between $P$, $Q$, $M$ and $N$ is:$P = \frac{N}{M}$ and $Q = \frac{N}{M} - 1 = \frac{{N - M}}{M}$. When the number of nodes in the cluster is relatively stable, it satisfies $P = E[X]$.

Since intra-cluster communication is managed by SCH based on TDMA, so there is no parallel transmission in the same time slot, the packet collision will not occur. Therefore, the intra-cluster throughput of a single cluster is
\begin{equation}
{S_{i1}} = \frac{{{P_{succ1}}QE[Packet1]}}{{{T_{e1}}}}.
\end{equation}
$P_{succ1}$, $E[Packet1]$ and $T_{e1}$ denote the probability of successful transmission, the average size of the intra-cluster packet per CM, and the frame period, respectively. Thus satisfying ${P_{succ1}} = {P_{res}}(1 - {P_{lost}})$ and ${P_{res}} = \frac{Q}{P} = \frac{{N - M}}{N}$, where $P_{res}$, $P_{lost}$ are probability of successful reservation and packet loss, respectively.

For PCH $i$, the number of messages that can be forwarded per frame is the minimum value of the total amount of messages sent by CMs to it and the frame length of the transmission period $T_s$, namely
\begin{equation}
E[Packet2] = \min \{ \sum\limits_i {{P_a},{T_s}} \},
\end{equation}
where $\sum\limits_i {{P_a} = {P_t}} {E_s}Q$. $E_s$ and $P_t$ are the average number of packets and the probability that a CM sends to PCH for inter-cluster communication, which satisfies $
{P_t} = \frac{{N - \frac{N}{M}}}{N} = \frac{{M - 1}}{M}.$

The inter-cluster throughput of a single cluster can be expressed as
\begin{equation}
\label{eq8}
\begin{aligned}
	{S_{i2}} &= \frac{{{P_s}{P_b}E[Packet2]}}{{{T_{e2}}}}
	{\kern 1pt} {\kern 1pt} {\kern 1pt} {\kern 1pt} {\kern 1pt} {\kern 1pt} {\kern 1pt} {\kern 1pt} {\kern 1pt} {\kern 1pt} {\kern 1pt} {\kern 1pt} {\kern 1pt} {\kern 1pt} \\
	& = \frac{{{P_s}{P_b}E[Packet2]}}{{(1 - {P_b}){T_{idle}} + {P_s}{P_b}{T_s} + (1 - {P_s}){P_b}{T_c}}},
\end{aligned}
\end{equation}
where $T_{e2}$, $T_{idle}$, $T_c$ and $T_s$ are the period, idle time, collision time and successful transmission time of each Markov state respectively.
$P_{tr}$ is denoted as the probability of inter-cluster communication between PCHs. Assume that the data arrival follows poisson distribution $\lambda_2$. The packets received by PCH is equivalent to the number of data needed to be forwarded between clusters by all nodes per frame, and it satisfies 
\begin{equation} 
{P_{tr}} = 1 - {e^{ - \lambda_2 {T_{e1}}}}.
\end{equation}

When the UAV is evenly distributed, the number of adjacent clusters also tends to be stable. Let the number of adjacent clusters be $R$, then the collision probability $P_c$ and the possibility of channel busy $P_b$ satisfies
\begin{equation}
\begin{aligned}
	\centering
	&{P_c} = 1 - {(1 - \tau )^{R - 1}}\\
	&{P_b} = 1 - {(1 - \tau )^R},
\end{aligned}
\end{equation}
the probability of successful forwarding $P_s$ satisfies 
\begin{equation}
{P_s} = \frac{{R\tau {{(1 - \tau )}^{R - 1}}}}{{{P_b}}}.
\end{equation}

B. The total throughput of the network

Assume that the structure of the network is grid. The network with 9 clusters is shown in Figure 5\subref{9c} below. When cluster $3$ is broadcasting to cluster $2$ and cluster $6$, cluster $4$ can also broadcast to cluster $1$, cluster $5$ and cluster $7$. Therefore, PCHs which are far away from each other can realize parallel broadcast in the same channel at the same time with little interference. By adopting CSMA, the throughput of the cluster-based network is different from that of the conventional single channel network. The key factor of the throughput of the cluster-based network is the number of PCHs which can broadcast parallelly. We give our solutions in the following discussion.
\begin{figure}[htbp]
\centering
\subfloat[9 clusters]{\label{9c}
	\includegraphics[width=0.2\textwidth]{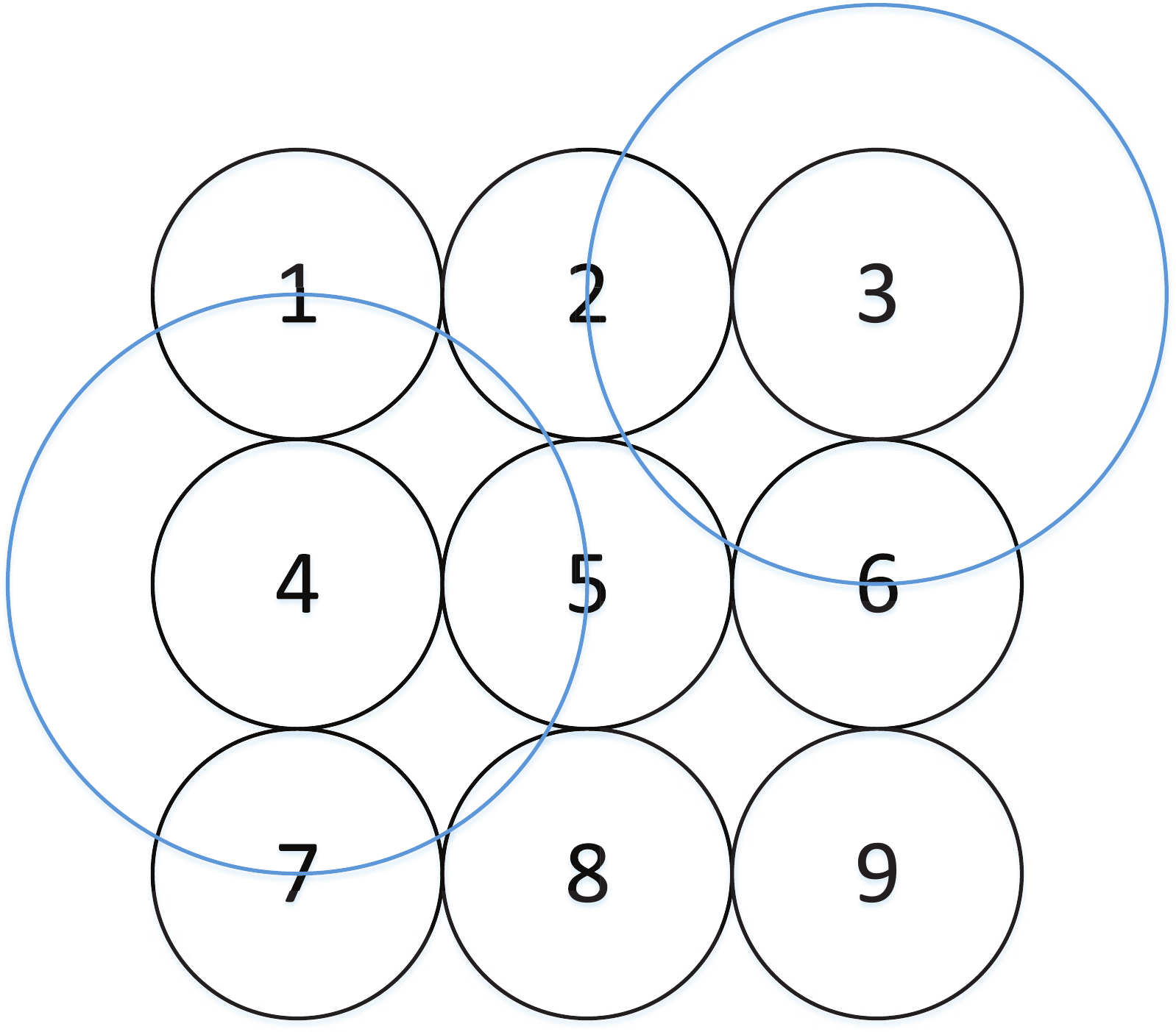}
}\hfill
\subfloat[clusters with $x$ = 4]{\label{x4}
	\includegraphics[width=0.2\textwidth]{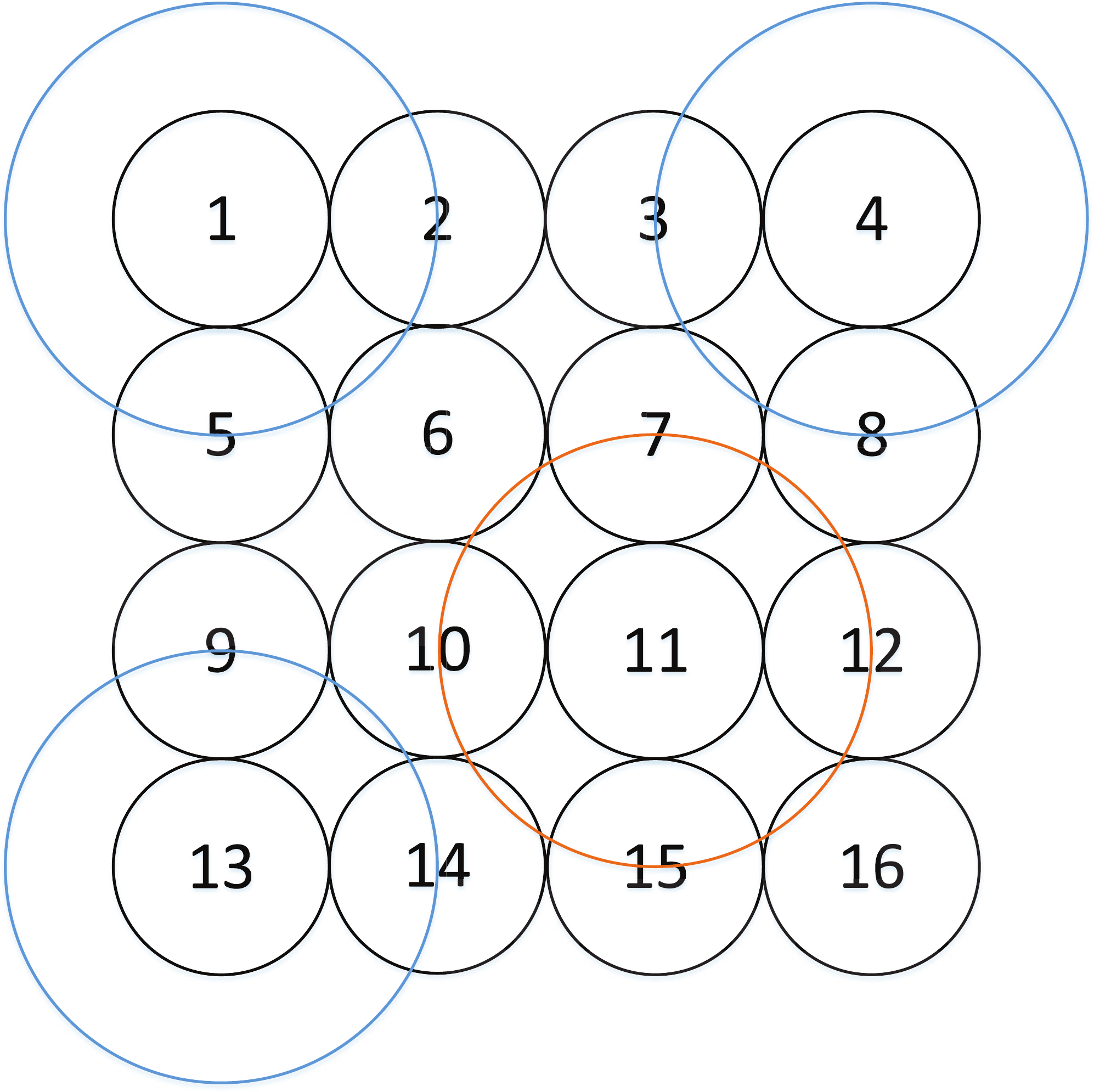}
}\newline
\subfloat[clusters with $x$ = 5]{\label{x5}
	\includegraphics[width=0.2\textwidth]{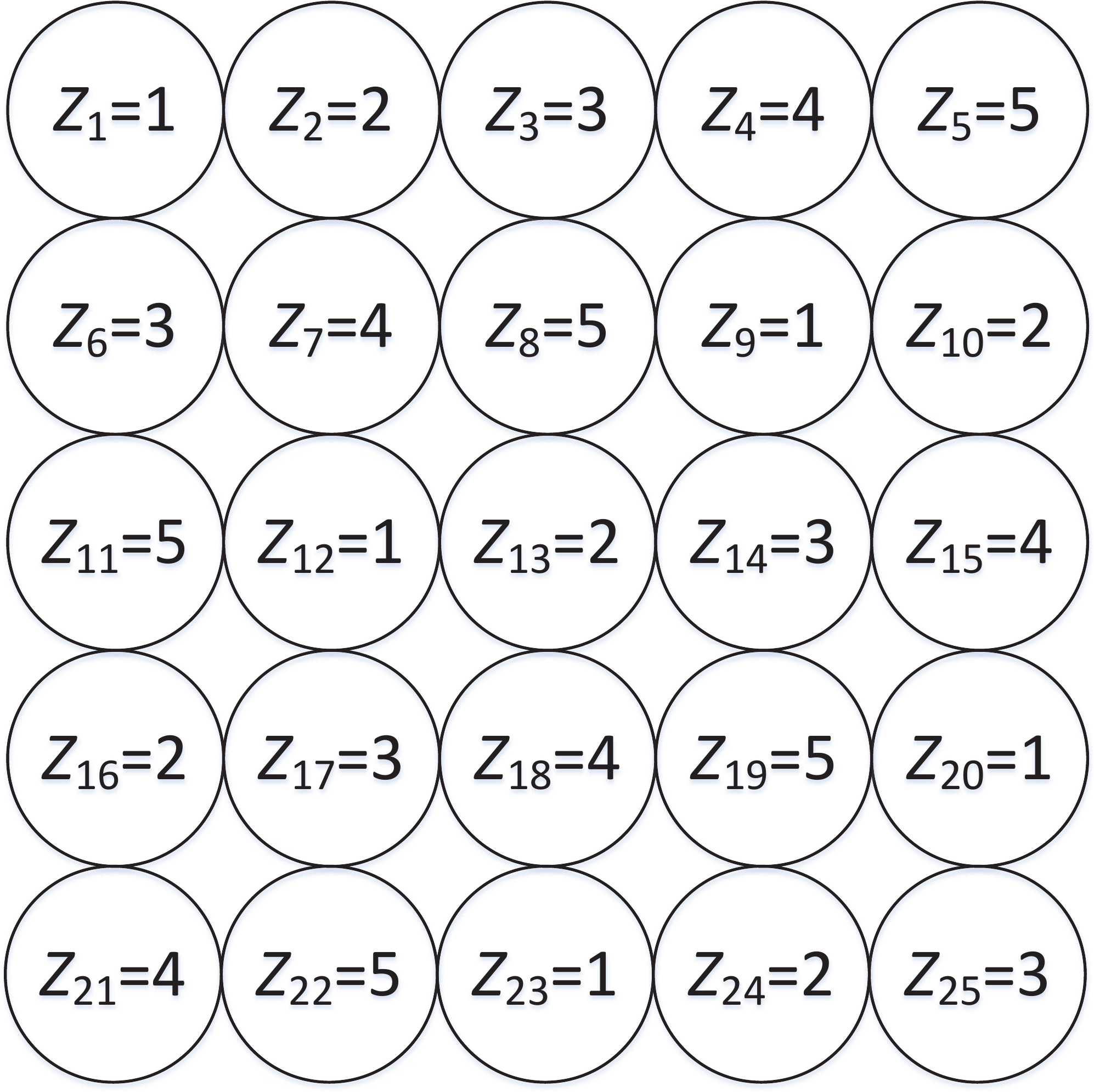}
}\hfill
\subfloat[clusters with $x$ = 6]{\label{x6}
	\includegraphics[width=0.2\textwidth]{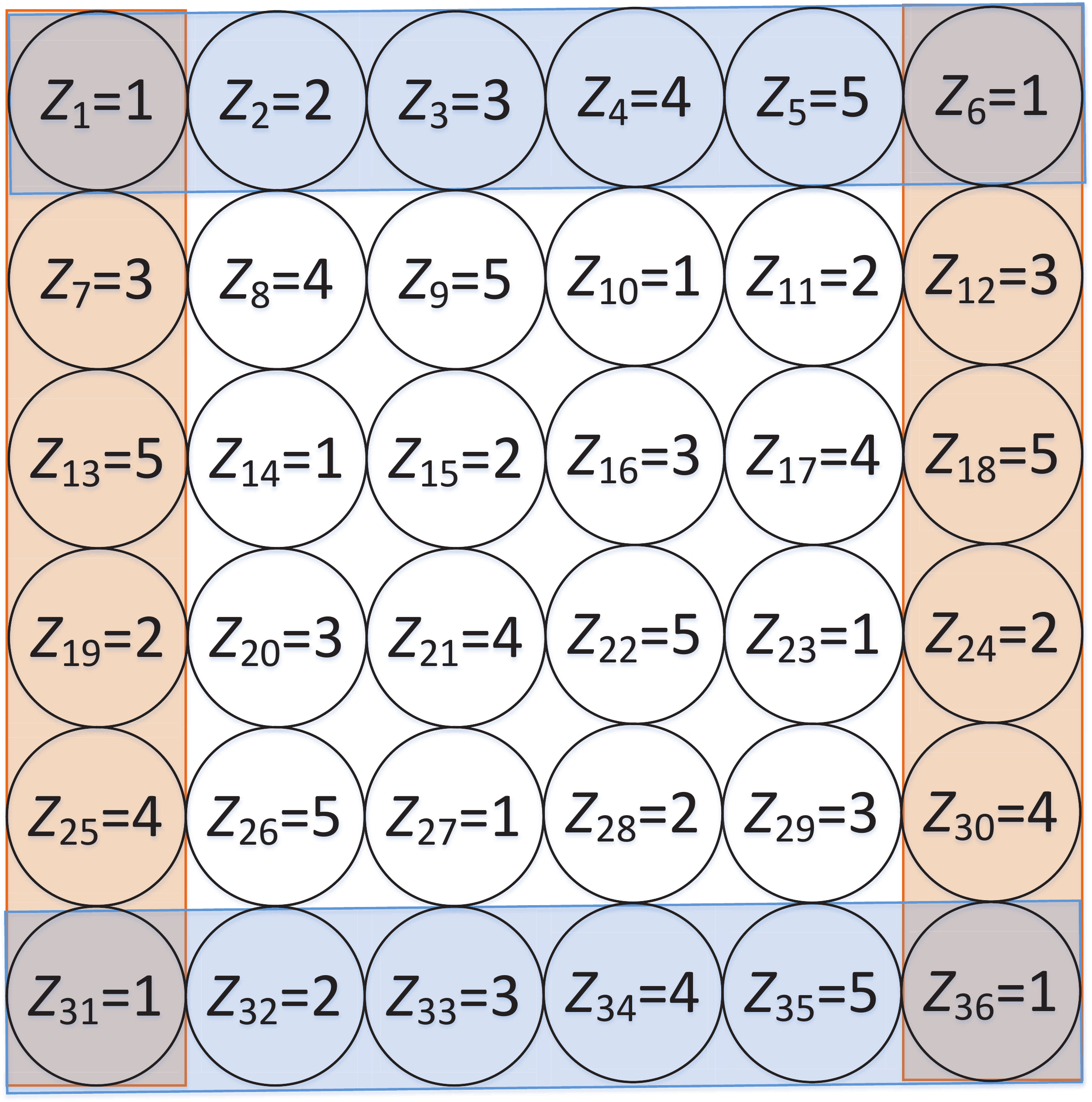}
}
\caption{Different number of clusters}
\label{fig5}
\end{figure}

The maximum number of clusters that can communicate at the same time, which is represented by $K$, is different and depends on the number of total clusters. Assume that there are $M$ clusters with $x$ rows and $x$ columns. The values of $K$ with different $x$ are discussed below:  

a) When $x=2$, $K=1$;  

b) When $x=3$, as shown in the Figure 5 \subref{9c} above, it is easy to know $K=2$;  

c) When $x=4$, as shown in the Figure 5 \subref{x4}, at most $K=4$ clusters can broadcast at the same time.  

d) When $x=5$, for the convenience of calculation, we consider another idea: the shortest successful broadcast period of each cluster. Assume that $T_x$  is the period it takes for a cluster to accomplish broadcast. The cluster has at least one chance to broadcast within $T$, and $T$ is $L$ times the length of $T_x$. The smaller $L$ is, the greater the chance of the inter-cluster broadcast is, and the higher the inter-cluster throughput is.  Then the problem becomes to find the minimum $L$ so that each cluster occupies at least one $T_x$ in $T$ to realize broadcast. Assume that clusters broadcast at specific intervals. The broadcast time of the $i$th cluster may be $l_i=1,2...,L$, so we can see that $L \ge 5$. When $L$ = 5, there is an allocation mode as shown in Figure 5\subref{x5}, so that each cluster can have a broadcast opportunity, where $Z_i$ represents the broadcast time interval of the $i$th cluster.  

e) When $x>5$, each cluster can use the same time point as the ($x-5$)th cluster to broadcast, Figure 5\subref{x6} shows the case of $x$= 6. To maximize throughput, we can know that $L$= 5 when $x \ge 5$, in this case we have $K={M}/{5}$.  

Hence, the value of $K$ is
\begin{equation}
K = \left\{ \begin{array}{l}
	1,{\kern 1pt} {\kern 1pt} {\kern 1pt} {\kern 1pt} x = 2;\\
	2,{\kern 1pt} {\kern 1pt} {\kern 1pt} x = 3;\\
	4,{\kern 1pt} {\kern 1pt} {\kern 1pt} x = 4;\\
	{M}/{5},{\kern 1pt} {\kern 1pt} {\kern 1pt} x \ge 5.
\end{array} \right.
\end{equation}

Suppose there are $M$ clusters in the network. For the $m$th cluster, the number of its neighbors is $R_m$, and it can realize parallel broadcasting with ($K_m-1$) clusters at most. Suppose that there are $k$ PCHs ($k \le K$) that simultaneously broadcast in a certain state, then the statistic value of inter-cluster throughput is
\begin{equation}
{S_M} = \sum\limits_{m = 1}^M {\left( {\sum\limits_{k = 1}^{{K_m}} {\left( {{{( - 1)}^{k + 1}}\sum\limits_{i = 1}^k {\frac{1}{{{R_i} + 1}}{S_{i2}}} } \right)} } \right)},
\end{equation}

which satisfies
\begin{equation}
s.t.\left\{ \begin{array}{l}
	{I_{ij}} > 2\\
	1 \le R_i \le 4\\
	{N_{{I_{ij}}}} \le 1.
\end{array} \right.
\end{equation}
${I_{i,j}} > 2$ represents that the distance between two clusters that simultaneously send messages should be more than twice the length of the broadcast radius. $1 \le {R_i} \le 4$ means the maximum number of adjacent clusters is four in the grid network. ${N_{{I_{ij}}}} \le 1$ represents that in all cases of simultaneous communication, duplication cannot occur. 

The total throughput of the network, which is composed of inter-cluster throughput and intra-cluster throughput, is expressed as
\begin{equation}
S = {S_M} + \sum\limits_{i = 1}^M {{S_{i1}}}.
\end{equation}

C. Energy consumption

Energy consumption is a major issue for wireless network lifetime. In a clustered network, the energy consumption of the CH will affect the overall lifetime of the network. The larger the amount of data, the greater the energy consumption \cite{24,25}. Therefore, we have
\begin{equation}
	{E_{consp}} = {E_{elec}} \times l,
\end{equation}
where ${E_{consp}}$ is the energy consumption, $E_{elec}$ represents the energy consumed when forwarding or receiving each data packet. $l$ is the total length of the data. 

The average energy consumed per cluster for one round is
\begin{equation}
	{E_r} = {E_{itr}} + {E_{ira}},
\end{equation}
where $E_{itr}$ is the inter-cluster energy consumption and $E_{ira}$ is the intra-cluster energy consumption. This consumption is calculated as the average energy consumed by the successful reception of the resulting packets, including the energy consumption of the retransmission. Retransmission happens when there is a packet loss or collision. So the probability of retransmission is equal to the probability of packet loss or collision. Since intra-cluster communication is based on TDMA, there is no packet collision. Hence $E_{ira}$ is expressed as

\begin{equation}
	\begin{aligned}
	&{E_{ira}} = E[Packet1] \times T{N_1} \times {E_{elec}}\\
	&T{N_1} = {P_{succ1}} \times {N_{s1}} \times {T_{s1}} + {P_{lost}} \times {N_{c1}} \times {T_{c1}}
	\end{aligned},
\end{equation}
where $N_{c1}$ and $N_{s1}$ are the number of loss packets and successful transmissions. $T_{c1}$ and $T_{s1}$ are the corresponding time duration. $E_{itr}$ is given as

\begin{equation}
	\begin{aligned}
	&{E_{itr}} = E[Packet2] \times T{N_2} \times {E_{elec}}\\
	&T{N_2} = {P_c} \times {N_{c2}} \times {T_{c2}} + {P_s} \times {N_{s2}} \times {T_{s2}}
	\end{aligned},
\end{equation}
where $N_{c2}$ and $N_{s2}$ are the number of data collisions and successful transmissions. $T_{c2}$ and $T_{s2}$ are the corresponding time duration.

As for the energy dissipated, we define it as $E_{diss}$, which is the differential value of the energy consumed to transmit a packet and energy consumed to successfully transmit the packet, satisfying

\begin{equation}
	{E_{diss}} = {E_r} - \left( {{S_{i1}} + {S_{i2}}} \right) \times {T_s}.
\end{equation}

\section{SIMULATION RESULTS}
\label{simulation results}
This section shows the simulation results of the proposed scheme compared with FM-MAC (flying ad hoc networks multi-channel MAC) \cite{12}. In order to ensure the same simulation conditions, the number of SCHs in FM-MAC is only two, and all packets have the same priority. 

\begin{figure}[htbp]
\centering
\includegraphics[width=0.45\textwidth]{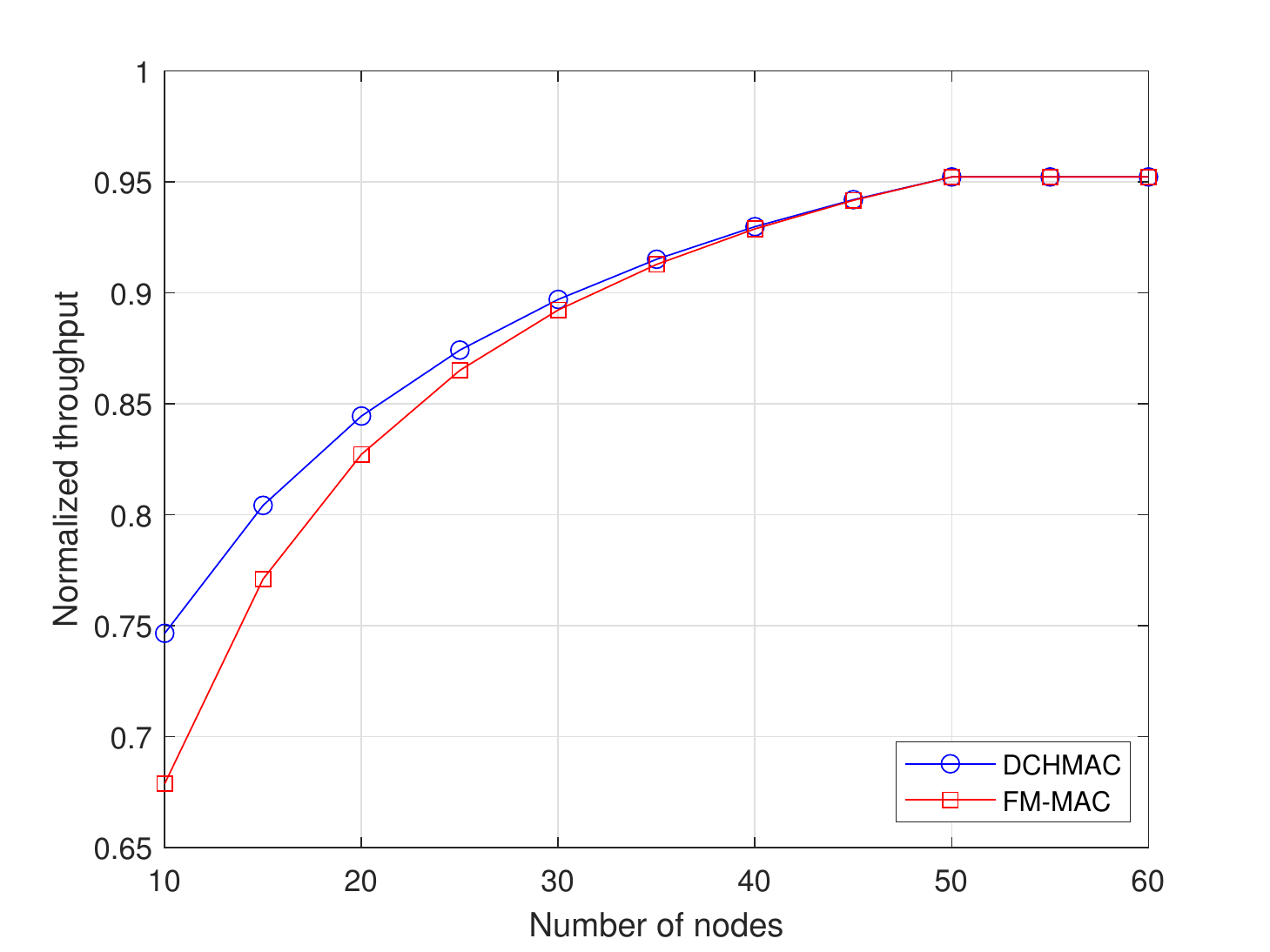}
\caption{Total throughput vs. the number of nodes within cluster.}
\label{fig6}
\end{figure}

Assume that the total number of UAVs is 200, and the threshold for the number of nodes in a cluster is 50. Figure~\ref{fig6} shows the impact of the number of nodes in the cluster on total normalized throughput. It can be seen that the throughput of DCHMAC scheme proposed in this paper is higher than that of FM-MAC. This is mainly because the dual cluster head scheme adopted by DCHMAC is favourable to the efficiency of communication. For example, it mitigates the load pressure of single CH and enables the parallel execution of intra-cluster communication and inter-cluster communication with less interference. 

\begin{figure}[htbp]
	\centering
	\includegraphics[width=0.45\textwidth]{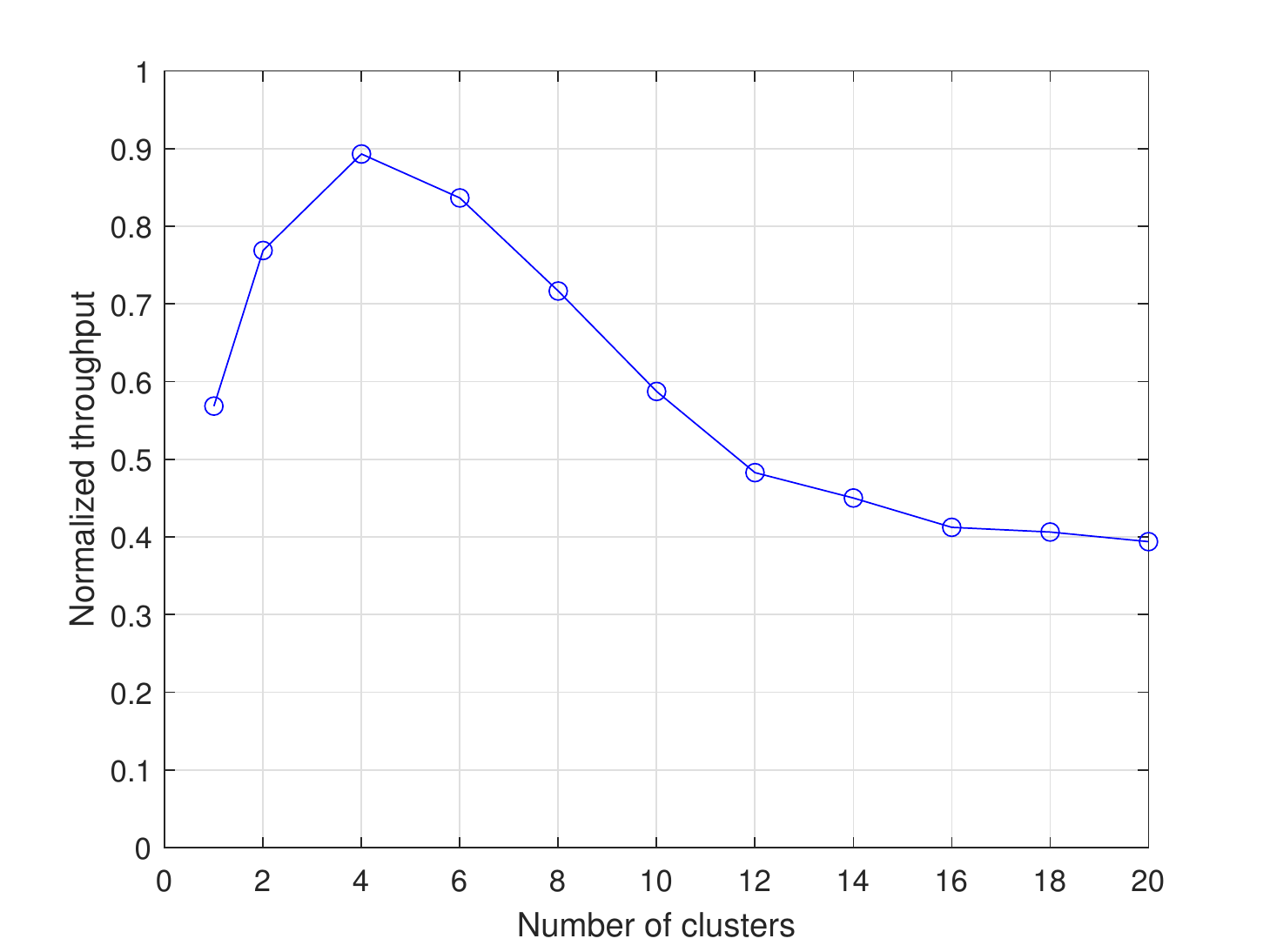}
	\caption{Total throughput vs. the number of clusters.}
	\label{fig7}
\end{figure}

When the total number of UAVs is certain, the number of clusters will affect the efficiency of intra-cluster communication and inter-cluster communication. Figure~\ref{fig7} shows the impact of the number of clusters on throughput. When the number of clusters increases gradually, the throughput decreases gradually. This is mainly because the probability of inter-cluster communication increases greatly when the number of clusters increases, leading to the rise of packet loss rate in inter-cluster communication and the rise of idle time slots in intra-cluster communication. The simulation result also gives an optional choice of selecting the number of clusters under the proposed scheme, which satisfies the need of throughput above 0.8. The favourable number is 3 to 6.   

\begin{figure}[htbp]
	\centering
	\includegraphics[width=0.45\textwidth]{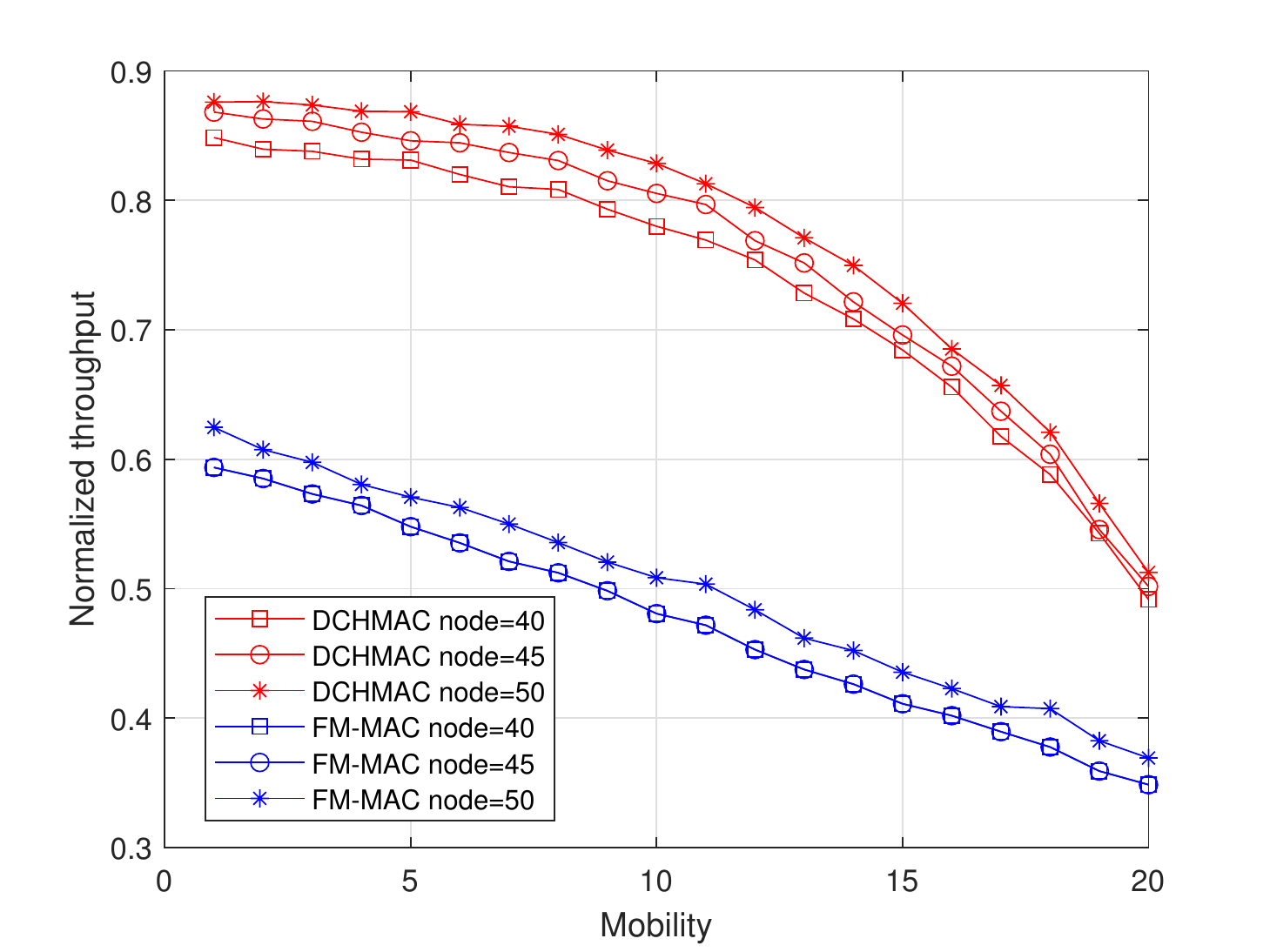}
	\caption{Total throughput vs. mobility.}
	\label{fig8}
\end{figure}

Figure~\ref{fig8} shows the impact of mobility on throughput, where Mobility represents the average number of UAVs leaving cluster per frame. It can be seen that with the increasing mobility of UAVs, the throughput of the two schemes will decrease continuously. While the throughput of DCHMAC maintains higher level than that of FM-MAC. This is because that by optimizing the frame structure, PCH can detect the status of UAVs more quickly, so the idle CIDs can be utilized in time. However, when the mobility increases to a certain extent, the detection ability of PCH can not keep up with the leaving rate of UAV, in which case the throughput will decrease greatly.

\begin{figure}[htbp]
\centering
\includegraphics[width=0.45\textwidth]{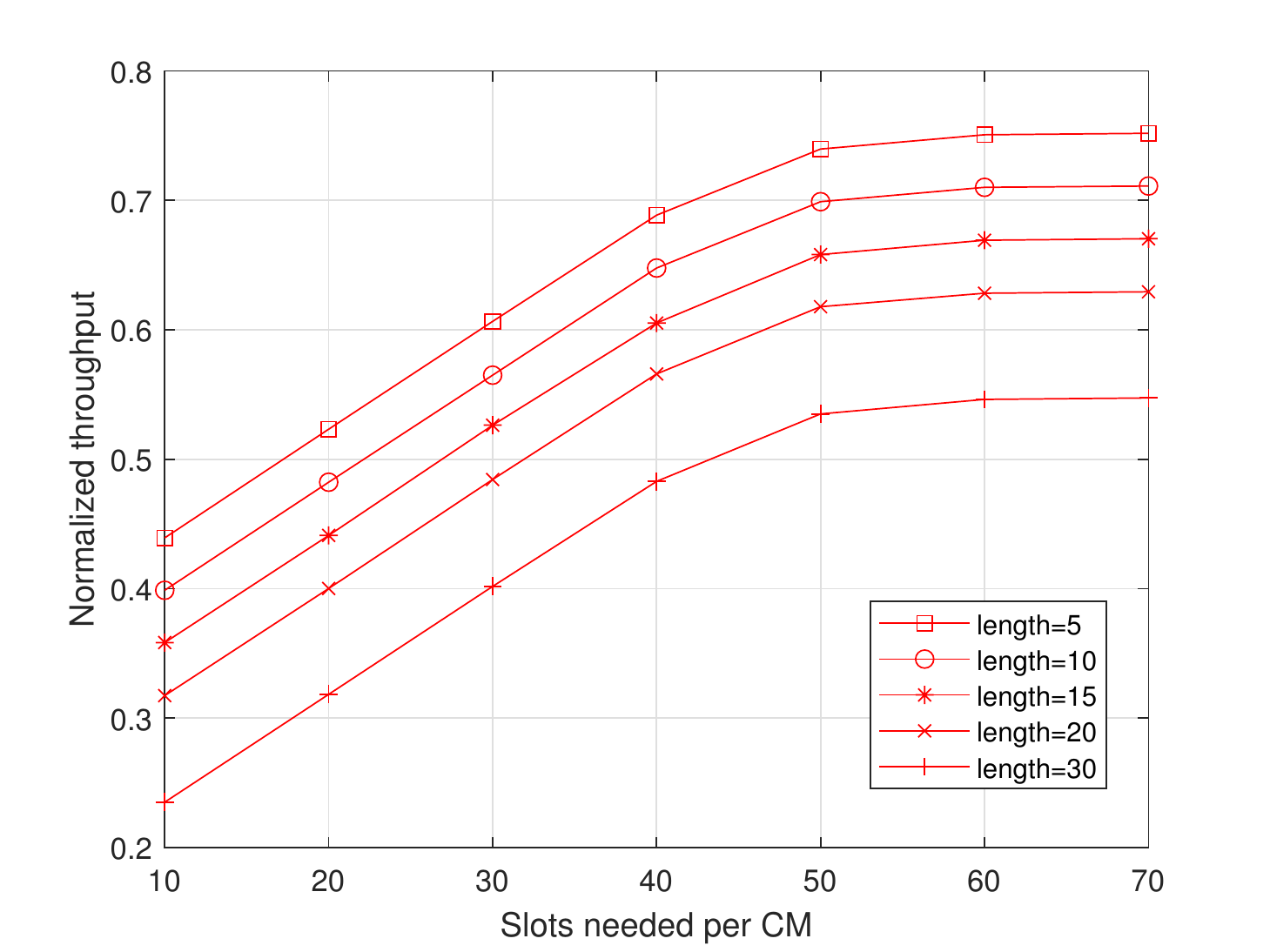}
\caption{Total throughput vs. the slots needed by CM.}
\label{fig9}
\end{figure}

Figure~\ref{fig9} shows the relationship between the time slots needed by CM per frame and the influence it has on throughput. In Figure~\ref{fig9}, length represents the length of the fixed period. We can see that it performs better when the CM applies for more time slots than the length of fixed time slots assigned for it.

\begin{figure}[htbp]
\centering
\includegraphics[width=0.45\textwidth]{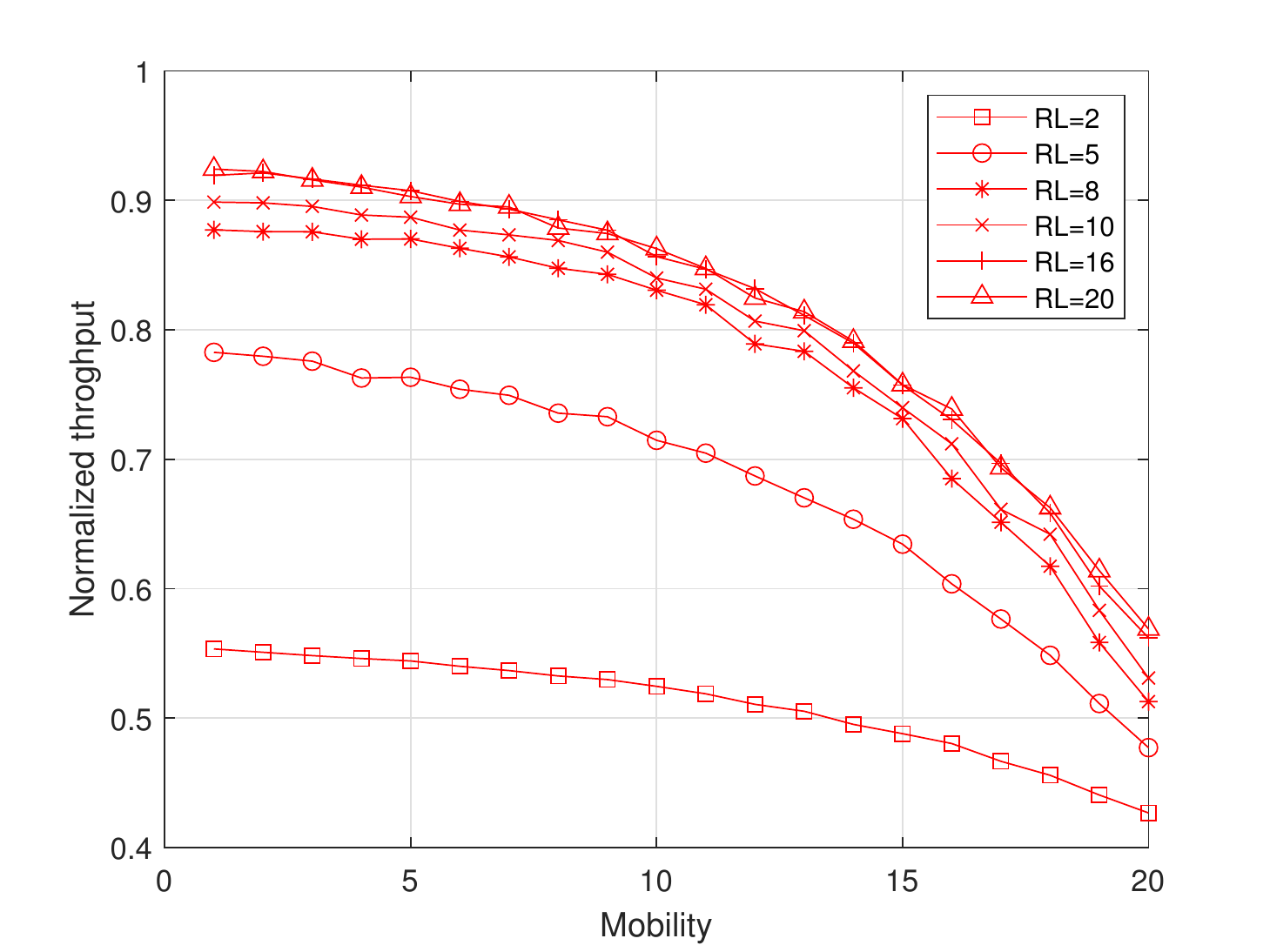}
\caption{Total throughput vs. Mobility.}
\label{fig10}
\end{figure}

Figure~\ref{fig10} further demonstrates the impact of the relative value of the time slots per CM on throughput, where RL is the ratio of the time slots needed by CM per frame to the length of fixed time slots. The figure shows that with the increase of mobility, there will be a decrease on performance, while with higher RL, the throughput can be much improved.

\begin{figure}[htbp]
	\centering
	\includegraphics[width=0.5\textwidth]{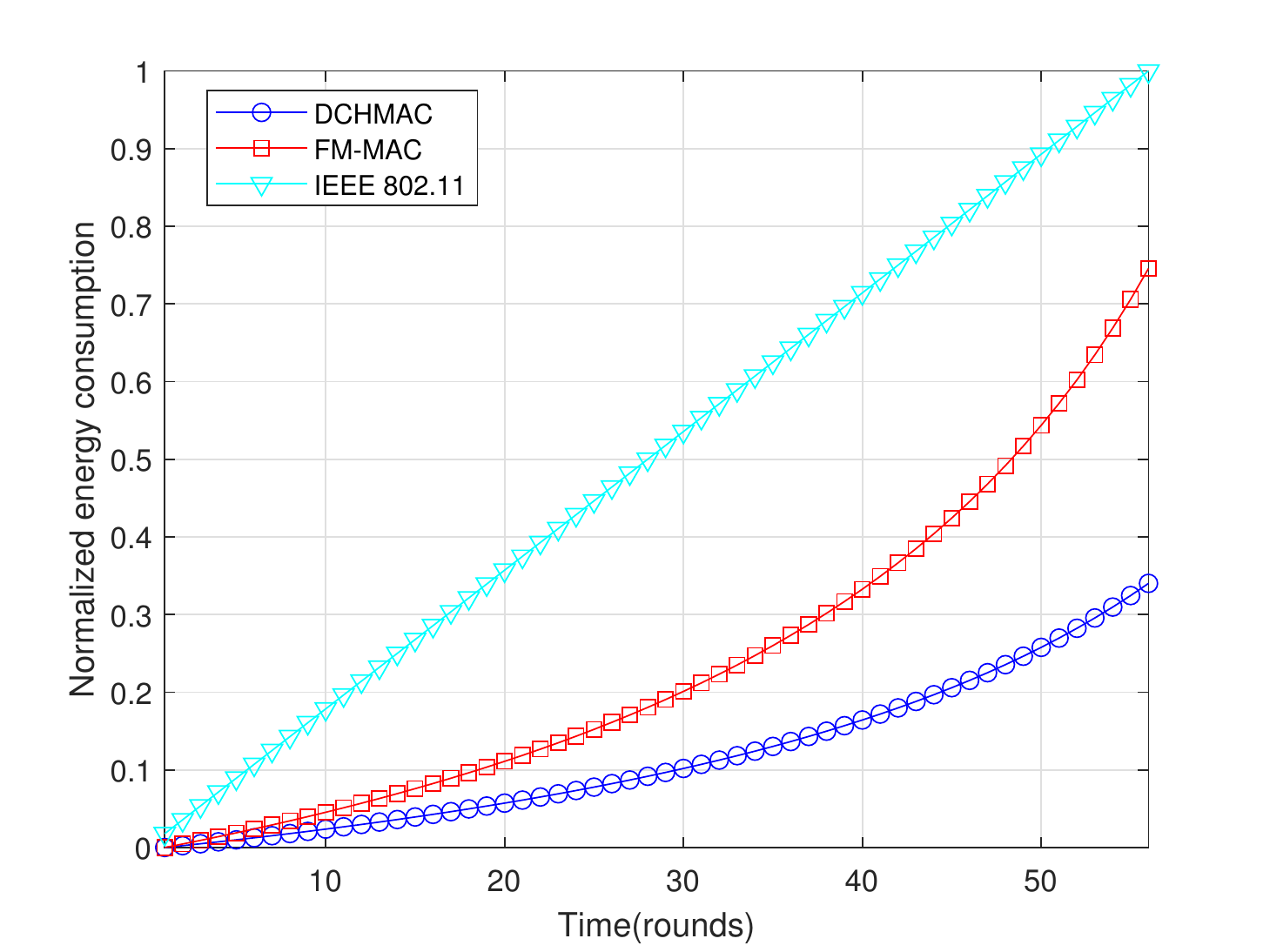}
	\caption{Energy consumption vs. Time.}
	\label{fig11} 
\end{figure}

To illustrate how the energy of CH is consumed over time, three mechanisms, namely IEEE 802.11, FM-MAC and DCHMAC, are adopted as shown in Figure~\ref{fig11}. Since the concept of CH does not exist in the IEEE 802.11, its energy consumption is represented by the average energy consumption of all nodes. In an attempt to focus on the trend, we normalized the energy consumption. This consumption is calculated as the average energy consumed by the successful reception of the resulting packets, including the energy consumption of the retransmission. As can be seen from Figure~\ref{fig11}, the energy consumption of the IEEE 802.11 scheme increases proportionally over time, while FM-MAC and DCHMAC have lower energy consumption, and DCHMAC has the lowest energy consumption. When the nodes in FM-MAC consume 0.74 of total energy, the nodes in DCHMAC only consumes around 0.34. Hence the proposed scheme can effectively prolong network lifetime about $40\%$ compared with FM-MAC. This is because DCHMAC can increase the access opportunity of nodes through the cooperation of dual CHs. Nodes send data to different CHs according to different communication requirements, instead of sending all the data to the same CH. In this way, the scheme of dual CHs can reduce the number of waits and conflicts of packets, thus reducing unnecessary energy consumption and leveraging network lifetime.

\begin{figure}[htbp]
	\centering
	\includegraphics[width=0.5\textwidth]{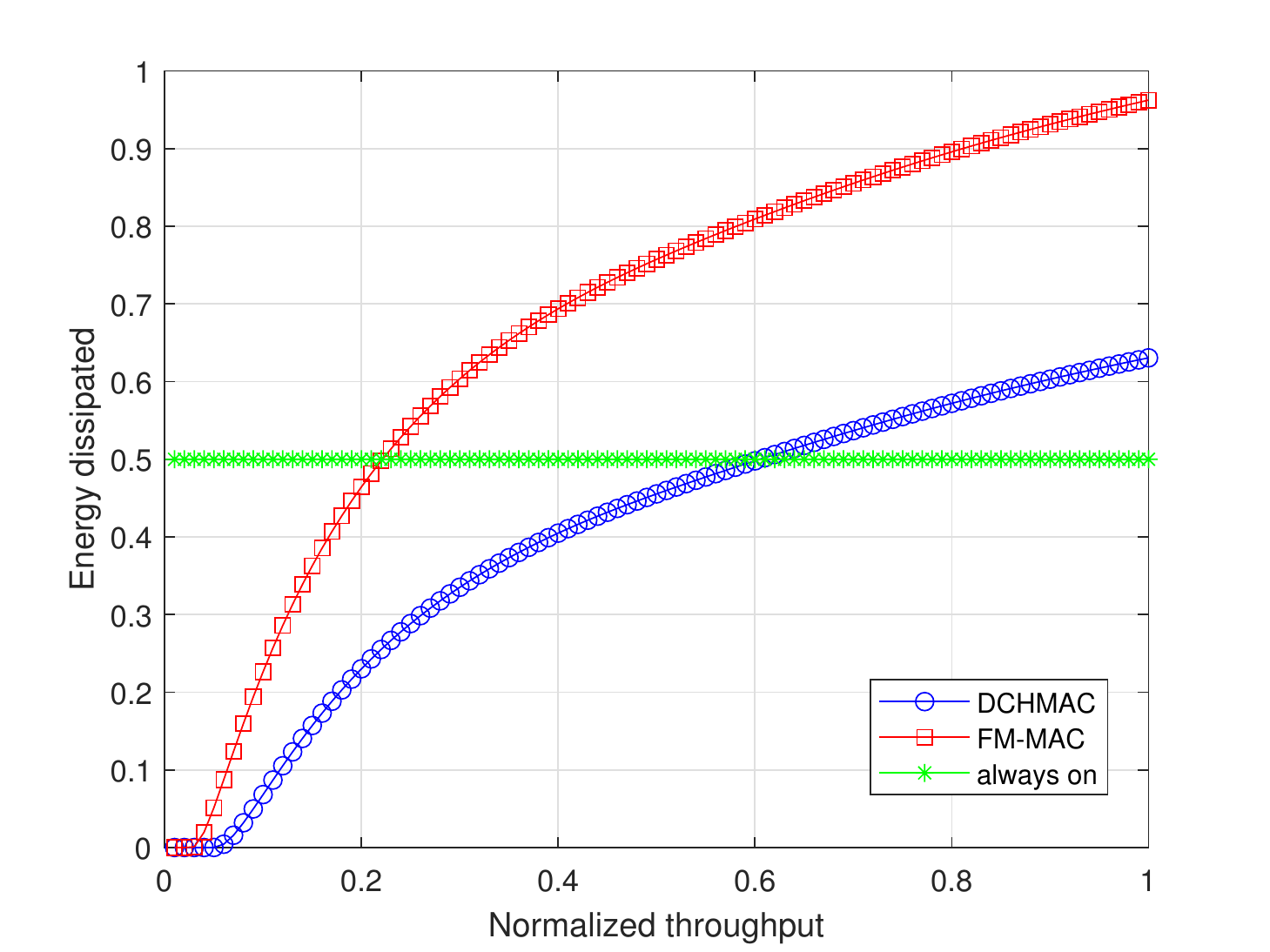}
	\caption{Dissipated energy vs. Desired throughput.}
	\label{fig12}
\end{figure}

To evaluate how increased throughput affects dissipation of energy, the dissipated energy in different throughput demand is performed, as shown in Figure~\ref{fig12}. Energy dissipated refers to the energy wasted for the desired throughput, i.e. the energy consumed by collisions, retransmissions, and discard of packets. Node can save energy by sleeping when not sending and receiving. At a low throughput demand, the node sleeps more, it wastes much less energy than the status of always on. In this case, the resource is redundant for the given throughput, so the energy dissipated for maintaining the given throughput is negligible. When the throughput demand continues to increase, the node is tend to continue to send and receive more data. Due to limited resources, the collision and packet loss rate will increase, leading to the increase of energy dissipation. In DCHMAC, through the collaboration and cooperation of the two CHs, the workload of a single CH can be effectively alleviated, and the process of intra-cluster communication and inter-cluster communication can be completed more efficiently. The energy dissipation in DCHMAC is 0.63 when the value of which is 0.96 in FM-MAC. So compared with FM-MAC, DCHMAC provides a lower amount of resource wastage, with a hike of $30\%$ energy efficiency.

\section{CONCLUSION}
\label{conclusion}
In this paper, a DCHMAC scheme for large-scale UAV ad hoc network is proposed, including UAV network clustering algorithm, channel division method and frame structure design. Different MAC schemes are adopted according to different communication scenarios, and the overload of a single CH is mitigated by the cooperation between PCH and SCH. Finally, theoretical analysis and simulations are discussed on the proposed scheme. The simulation results show that DCHMAC improves the throughput of the network, reduces the energy dissipation of the network, and prolongs the network lifetime.

\bibliographystyle{gbt7714-numerical}
\bibliography{zxr_2}

\end{document}